\documentclass[structabstract]{aa}  
\usepackage{graphicx}
\usepackage{txfonts}
\begin{document}
   \title{High-energy radiation from the massive binary system Eta Carinae}


   \author{W. Bednarek 
          \and
          J. Pabich
          }

   \institute{Department of Astrophysics, University of \L \'od\'z,
              ul. Pomorska 149/153, 90-236 \L \'od\'z, Poland\\
              \email{bednar@astro.phys.uni.lodz.pl; jpabich@uni.lodz.pl}
             }

   \date{Received ; accepted }


\abstract
{The most massive binary system Eta Carinae has been recently established as a gamma-ray source by the AGILE and Fermi-LAT detectors. The high energy spectrum of this gamma-ray source is very intriguing. It shows two clear components and a lack of any evidence of variability with the orbital period of the binary system.}
{We consider different scenarios for the acceleration of particles (both electrons and hadrons) and the production of the high energy radiation in the model of stellar wind collisions within the binary system Eta Carinae with the aim to explain the gamma-ray observations and predict the behaviour of the source at very high gamma-ray energies.}
{The gamma-ray spectra calculated in terms of the specific models are compared with the observations of Eta Carinae, and the neutrino spectra produced in hadronic models are confronted with the atmospheric neutrino background and the sensitivity of 1 km$^2$ neutrino telescope.}
{We show that spectral features can be explained in terms of the stellar wind collision model between the winds of the companion stars in which the acceleration of particles occurs on both sides of the double shock structure. The shocks from the Eta Carinae star and the companion star can accelerate particles to different energies depending on the different conditions determined by the parameters of the stars. The lack of strong GeV gamma-ray variability with the period of the binary system can be also understood in terms of such a model.}
{We predict that the gamma-ray emission features at energies above $\sim$100 GeV will show significant variability (or its lack) depending on the acceleration and interaction scenario of particles accelerated within the binary system. For the hadronic models we predict the expected range of neutrino fluxes from the binary system Eta Carinae. This can be tested through observations with the large-scale neutrino telescopes, which will support or disprove the specific hadronic models.}

\maketitle
%

%
%
\section{Introduction}

Eta Carinae is the highest mass binary system detected up to now. It contains
a star with a mass estimated in the range $80-120$M$_\odot$ (Davidson \& Humphreys~1997, Hillier et al.~2001) up to $200$M$_\odot$ (Kashi \& Soker~2010).
The basic parameters of this star remain uncertain. Its luminosity is equal to $4.5\times 10^6$L$_\odot$, and the surface temperature to $2\times 10^4$ K. The radius estimates of Eta Carinae can differ by a factor of a few, between $40-180$ R$_\odot$ (Hillier et al.~2001). The star creates a very strong wind with a terminal velocity of $\sim 500-700$ km s$^{-1}$. The mass loss rate is estimated to be in the range from $2.5\times 10^{-4}$ M$_\odot$ yr$^{-1}$ (Pittard \& Corcoran~2002) up to $10^{-3}$ M$_\odot$ yr$^{-1}$ (Hillier et al.~2001). Moreover, this mass loss rate seems to change rapidly (Martin et al.~2010, Corcoran et al.~2010).
Eta Carinae is suspected to be a binary system with a period of $\sim 2022.7\pm 1.3$ days (Daminelli et al.~2008), characterised by high eccentricity $e\sim 0.9$ (Nielsen et al.~2007) and the semi-major axis  of $a = 16.64$ AU (Hillier et al.~2001).  
The parameters of the companion star are even more uncertain. It is a WR type or an O type supergiant with a luminosity $9\times 10^5$L$_\odot$, and a surface temperature $\sim 4\times 10^4$ K (Verner, Bruhweiler \& Gull~2005, Mehler et al.~2010). The radius of the companion star estimated for the above parameters is $\sim 1.4\times 10^{12}$ cm.
The mass loss rate of the companion star is estimated to be ${\dot M}_{\rm comp}\sim 10^{-5}$ M$_{\odot}$/yr and the wind velocity to be $\sim 3000$ km s$^{-1}$ (Pittard \& Corcoran~(2002).
The timing of the periastron passage is defined as $T_0 = JD 245 2819.8$ (Damineli et al.~2008).
The binary system is immersed in the massive nebula ($M_{\rm neb}\sim 12$ M$_{\odot}$, Smith et al.~2003), which likely originated in the huge outburst observed in this system in the year 1843. The radius of the nebula is estimated to be $3.4\times 10^{17}$ cm (Smith et al.~1998).

The GeV $\gamma$-ray emission has been recently reported from the direction of Eta Carinae by the AGILE telescope (Tavani et al.~2009). The $\gamma$-ray flux was steady during the period of over one year corresponding to the pre-periastron passage except for a single flare lasting for about two days. The analysis of the whole Fermi-LAT data shows the $\gamma$-ray source to be consistent with the AGILE discovery (Abdo et al.~2010, Walter, Farnier \& Leyder~2010). These observations, covering also the period of the periastron passage, suggest a steady nature of the source. The $\gamma$-ray spectrum extends up to $\sim$100 GeV, showing two distinct components. The first one is consistent with a power law spectrum (spectral index $1.6\pm 0.2$) and a cut-off at  $1.6$ GeV. The second one is described by a power law with spectral index $\sim$1.9 without evidence of a cut-off.
The non-thermal nature of the source is also confirmed by the hard X-ray observations with the Beppo-Sax (Viotti et al.~2004), the INTEGRAL (Leyder et al.~2008) and the SUZAKU (Sekiguchi et al.~2009) satellites. This hard X-ray emission creates an additional flat component above the extended soft thermal X-rays which dominate below $1.5$ keV.
The X-ray emission from Eta Carinae in the 22-100 keV energy range is very hard (differential spectral index equal to $\cong 1\pm 0.4$, Leyder et al.~2008). More recent observations with INTEGRAL show that the hard X-ray component is well described by a power law spectrum (with photon index 1.8) without any strong variability at the periastron passage (Leyder et al.~2010). 

In general, non-thermal high-energy radiation from massive binary systems is interpreted as radiation from particles accelerated at the stellar wind shock (e.g. Eichler \& Usov~1993, Benaglia \& Romero~2003, Bednarek~2005, Reimer et al.~2006, Pittard \& Dougherty~2006 or Walter, Farnier \& Leyder~2010). Both, leptons and hadrons can contribute to the $\gamma$-ray emission from the binary systems in the considered scenarios. 
Particles might also be accelerated in the large-scale blast wave produced during the Eta Carinae explosion in the year 1843 (e.g. Ohm, Hinton \& Domainko~2010). It is argued that accelerated hadrons cannot find enough targets for efficient $\gamma$-ray production. However, accelerated electrons are able to produce GeV $\gamma$-rays in the inverse-Compton process by scattering infrared radiation from the nebula.

In this paper we analyse different scenarios for the high-energy production in the general wind shock model with an application to the most massive compact binary system Eta Carinae. In contrast to previous works, we consider the acceleration of particles in two shocks with different properties appearing from the side of the Eta Carinae star and the companion star.

\section{General scenario}

We consider a simple scenario for the high-energy production within the binary system of two massive stars similar to what was already discussed in the papers mentioned above. The massive stars produce very strong winds, which collide within the binary system and create a double shock structure (see Fig.~1). The location of the shock within the binary system, $R_{\rm sh}$, can be estimated for the known value of the coefficient $\eta = {\dot M_{\rm comp}}v_{\rm comp}/({\dot M_{\rm EC}}v_{\rm EC})$, which describes the pressure ratio of the winds from both stars. For the Eta Carinae binary system the wind pressure from the side of the Eta Carinae dominates. The value of $\eta$ is estimated to be $\sim 0.2$ by Pittard \& Corcoran~(2002) for the following parameters of Eta Carinae, $v_{\rm EC}\sim$500-700 km s$^{-1}$ and 
${\dot M}_{\rm EC}\sim 2.5\times 10^{-4}$ M$_\odot$ yr$^{-1}$, and the companion star, ${\dot M}_{\rm comp}\sim 10^{-5}$ M$_{\odot}$/yr and the wind velocity on $v_{\rm comp}\sim 3000$ km s$^{-1}$. This means that the shock is bound around the companion star but it is closer to the surface of the Eta Carinae star than to the surface of the companion star considering their stellar radii.
In the example modelling we present, we apply the above parameters for the stars and their stellar radii equal to $R_{\rm EC} = 1.2\times 10^{13}$ cm and $R_{\rm comp} = 1.4\times 10^{12}$ cm .
The distance of the shock from the companion star can be obtained from $R_{\rm sh}^{\rm comp} = D\sqrt{\eta}/(1 + \sqrt{\eta})$ and from Eta Carinae $R_{\rm sh}^{\rm Eta} = D/(1 + \sqrt{\eta})$, where $D$ is the separation of the stars. For the periastron passage, $D = (1 - e)a\approx 1.7$ AU. Then, $R_{\rm sh}^{\rm comp}\approx 6R_{\rm comp}\approx 8.4\times 10^{12}$ cm and $R_{\rm sh}^{\rm Eta}\approx 1.4R_{\rm Eta}\approx 1.7\times 10^{13}$ cm. For the apastron passage, the shock is located
about an order of magnitude farther from both stars owing to the highly eccentric orbit. However, the fractions of their wind powers, which are intercepted by the shock structure, are comparable.
The above mentioned estimates are obtained on the base of a simplified scenario. They may not correspond to the real situation as it occurrs in very close binary systems. At first, the stellar winds accelerate from the stellar surface. They are slower close to the stellar surfaces and reach the terminal speeds only at some distance from the stars. This may influence the localization of the shock within the binary system. Secondly, the radii of the companion stars are not preciously known. This means that in reality the distances of the acceleration region from the stellar surfaces may differ from the values estimated above. Thirdly, at the periastron passage the winds may not even balance. Owing to the instabilities in the wind, the shock surface can obtain a very complicated structure or it may even collapse onto one of the stars (e.g. Parkin et al.~2011). Therefore, the above estimated values should be taken with caution. They are applied as example values, but whether they correspond to the real situation in the Eta Carinae binary system is uncertain.

It is believed that shocks produced in collisions of stellar winds are able to accelerate particles to relativistic energies. Note, however, that the conditions at the shocks from the sides of both stars can be quite different. The mechanical pressure of the winds has to balance at the shock, i.e., parts of the power provided to the shock region from the sides of both stars are comparable. However, other physical parameters such as the magnetic field strength, the density of the wind or the wind velocity, can differ significantly. Therefore, it is expected that spectra of leptons and hadrons accelerated at the shocks from the sides of both stars can have different properties because of the  differences in the efficiency of the acceleration process, magnetic field strengths and energy losses of particles that can determine their maximum energies. The density of stellar photons in both shocks is expected to be similar owing to relatively small thickness of the double shock structure. 

In principle both shocks can accelerate leptons and hadrons to different maximum energies. Particles accelerated in different shocks can produce radiation at different energy ranges and at different dominant radiation processes. 
We also expect that depending on the shock (from the Eta Carinae or the companion star), particles can already lose energy close to the shock (i.e. within the binary system) or escape to the surrounding large-scale nebula. Therefore, high-energy radiation with very different properties can be expected in this relatively simple scenario of the two stellar wind interactions. 

\begin{figure}
\vskip 7.2truecm
\includegraphics{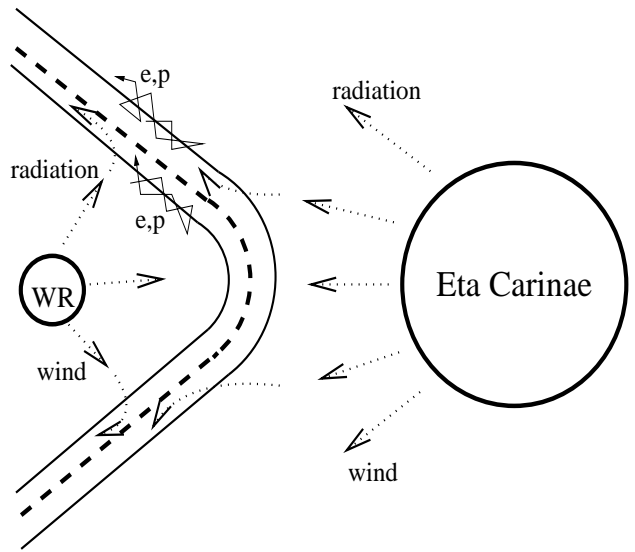}
\caption{Schematic representation of the supermassive binary system Eta Carinae. Electrons (e) and hadrons (p) can be accelerated on both shocks to different maximum energies due to different conditions at the shocks from the sides of the Eta Carinae star and the companion star. The details of the considered scenario are given in the main text of the paper.}
\label{fig1}
\end{figure}
\section{Acceleration of particles}

In general, the acceleration rate of particles at the shock can be parametrised by
\begin{eqnarray}
{\dot P}_{\rm acc} = \xi cE/R_{\rm L}\approx 0.1\xi_{-5} B~~~{\rm GeV~s^{-1}},
\label{eq1}
\end{eqnarray}
where $\xi = 10^{-5}\xi_{-5}$ is the acceleration parameter, $E$ is the energy of particle (in GeV), $R_{\rm L}$ is the Larmor radius of particles in the magnetic field $B$ (in Gauss), and $c$ is the velocity of light. We can estimate the characteristic acceleration time scale of particles from
\begin{eqnarray}
\tau_{\rm acc} = E/{\dot P}_{\rm acc}\approx 10E/(\xi_{-5}B)~~~{\rm s}.
\label{eq2}
\end{eqnarray}

This acceleration process can be limited either by the advection of particles along the shock surfaces with the stellar winds or by their energy losses through different radiation processes. The characteristic time scale for the advection process along the shock is, 
\begin{eqnarray}
\tau_{\rm adv} = 3R_{\rm sh}/v_{\rm w}\approx 3\times 10^5R_{13}/v_3~~~{\rm s},
\label{eq3}
\end{eqnarray}
where $R_{\rm sh} = 10^{13}R_{13}$ cm is the distance of the shock from the centre of the star and $v_{\rm w} = 10^8v_3$ cm s$^{-1}$ is the stellar wind velocity. 
The radiation energy losses of particles can determine their maximum energies. Below we estimate the energy loss time scales for electrons and hadrons and their maximum allowed energies at the shocks from the sides of the Eta Carinae and the companion star.

\subsection{Electrons}

The synchrotron and ICS processes for the electrons are important.  
The cooling time scale of electrons on the synchrotron and IC (in the Thomson regime) processes can be estimated from,
$\tau_{\rm syn/IC} = E_{\rm e}/{\dot P}_{\rm syn/IC}$, where $E_{\rm e}$ is the electron energy and ${\dot P}_{\rm syn/IC}$ are the energy loss rates of electrons on both processes. These time scales are estimated as,
\begin{eqnarray}
\tau_{\rm syn} = {{E_{\rm e}m_{\rm e}^2}\over{4/3c\sigma_{\rm T}U_{\rm B}E_{\rm e}^2}}\approx {{3.7\times 10^5}\over{B^2E_{e}}}~~~{\rm s},
\label{eq4}
\end{eqnarray}
\noindent
and
\begin{eqnarray}
\tau_{\rm IC/T} = {{E_{\rm e}m_{\rm e}^2}\over{4/3c\sigma_{\rm T}U_{\rm rad}E_{\rm e}^2}}\approx {{170}\over{E_{\rm e}}}\left[\left({{T_4^4}\over{R_{\rm sh}^2}}\right)_{\rm comp} + \left({{T_4^4}\over{R_{\rm sh}^2}}\right)_{\rm EC}\right]^{-1}~~~{\rm s},
\label{eq5}
\end{eqnarray}
\noindent
where $\sigma_{\rm T}$ is the Thomson (T) cross section, $U_{\rm B}$ and $U_{\rm rad}$ is the energy density of the magnetic and radiation fields, $m_{\rm e}$ is the electron mass, $R_{\rm sh}$ is the distance of the shock from a specific star expressed in units of the stellar radius of the specific star, and $T = 10^4T_4$K is the surface temperature of the specific star. The cooling time scale in the Klein-Nishina (KN) regime can be roughly estimated by introducing the energy of electrons into the above formula for the IC losses, corresponding to the transition between the Thomson and the Klein-Nishina regimes,
$E_{\rm e}^{\rm T/KN} = m_{\rm e}^2/(3k_{\rm B}T)\approx 97/T_4$ GeV. 
Then, the cooling time scale of electrons in the KN regime in the radiation field of both stars is approximately given by
\begin{eqnarray}
\tau_{\rm IC}^{\rm KN} = {{3E_{\rm e}m_{\rm e}^2}\over{4c\sigma_{\rm T}U_{\rm rad}(E_{\rm e}^{\rm T/KN})^2}}\approx 0.27E_{\rm e}\left[\left({{T_4^4}\over{R_{\rm sh}^2}}\right)_{\rm comp} + \left({{T_4^4}\over{R_{\rm sh}^2}}\right)_{\rm EC}\right]^{-1}~{\rm s}.
\label{eq6}
\end{eqnarray}
\noindent
The cooling time scale of electrons on the bremsstrahlung process is almost independent of the electron energy. It can be estimated from 
\begin{eqnarray}
\tau_{\rm br}\approx X_{\rm o}/c\approx 4.3\times 10^4R_{13}^2v_3/M_{-4}~~~{\rm  s}, 
\label{eq7}
\end{eqnarray}
\noindent
where $X_{\rm o} = 3.9\times 10^{25}/\rho_{\rm w}$ cm is the electron radiation length in hydrogen, $\rho_{\rm w} = {\dot M}/(4\pi R_{\rm sh}^2v_{\rm w})\approx 3\times 10^{10}M_{-4}R_{13}^{-2}v_3^{-1}$ cm$^{-3}$ is the concentration of matter in the stellar wind, and the mass loss rate of the star is ${\dot M} = 10^{-4}M_{-4}$ M$_\odot$ yr$^{-1}$. 

The synchrotron and IC cooling time scales are clearly shorter than the advection time scale of electrons along the shock and their bremsstrahlung energy loss time scale for the applied parameters of the stars in the binary system. Therefore, the maximum energies of accelerated electrons should be determined by balancing the energy gains from the acceleration process with the energy losses on the synchrotron and IC processes. The efficiency of these two processes depends on the energy density of the magnetic and radiation fields in the 
acceleration region. The synchrotron losses dominate over the IC losses in the T regime for the magnetic field
$B_{\rm sh} > 40[(T_4^4/R_{\rm sh}^2)_{\rm comp} + (T_4^4/R_{\rm sh}^2)_{\rm EC}]^{1/2}$ G. For the part of the shock that is the closest to the stars at the periastron passage, this limiting value is  $B_{\rm sh} > 160$ G. Depending on the specific parameters of the companion stars, we expect a variety of radiation scenarios from the binary systems:

\begin{enumerate}

\item the synchrotron energy losses always dominate over the IC process, 

\item the synchrotron energy losses dominate only at
the highest energies (whereas the IC scattering occurs in the Klein-Nishina regime),

\item the IC energy losses dominate for the whole energy range of accelerated electrons.

\end{enumerate}

\noindent
These radiation processes can occur in a different way on both sides of the double shock structure.

By balancing the synchrotron energy losses with the acceleration energy gains, we obtain the limit of the maximum energies of accelerated electrons at the shock,
\begin{eqnarray}E_{\rm e}^{\rm max}\approx 190(\xi_{-5}/B)^{1/2}~~~{\rm GeV}.
\label{eq8}
\end{eqnarray}
\noindent
On the other hand, by balancing the acceleration time scale with the IC energy loss time scale (in the T regime), we estimate the maximum energies of electrons as
\begin{eqnarray}
E_{\rm e}^{\rm max}\approx 4(\xi_{-5}B_{\rm sh})^{1/2}\left[\left(T_4^4/R_{\rm sh}^2\right)_{\rm comp} + 
\left(T_4^4/R_{\rm sh}^2\right)_{\rm EC}\right]^{-1/2}~~~{\rm GeV}.
\label{eq9}
\end{eqnarray}

To consider the high-energy processes in the Eta Carinae binary system, we have to fix the basic parameter that describes the considered scenario, which is the strength of the surface magnetic field of both stars. As an example, we use the value of $2\times 10^3$ G for the surface magnetic field of the companion star and $200$ G for
the Eta Carinae star. Then, the magnetic field at the shock from the companion star is estimated to be $B_{\rm sh}^{\rm comp}\sim 60$ G and at the shock from the Eta Carinae star as $B_{\rm sh}^{\rm comp}\sim 100$ G. We also fixed the values of the acceleration parameter $\xi$ for both shocks by estimating them from
$\xi\sim (v_{\rm w}/c)^2$. Because the velocity of the winds from both stars
differs significantly, we obtain, $\xi_{\rm EC}\sim 5\times 10^{-6}$ and $\xi_{\rm comp}\sim 10^{-4}$. Because of large differences between these acceleration coefficients and magnetic field strengths at the shocks, the
maximum energies of the electrons accelerated on both shocks (from the sides of Eta Carinae and the companion star) differ significantly.

The magnetic field strength in the winds of stars has a complicated dependence on the distance from the star. In a small region close to the stellar surface 
the magnetic field has a dipole structure, i.e., $B(R)\propto R^{-3}$ (USO \& Melrose~1992). However, in most cases this region is very small and it can be 
neglected. At larger distances the magnetic field has a radial structure. Then, its strength drops as $B(R)\propto R^{-2}$. This dependence dominates for distances characteristic for the periastron and apastron passages of the stars in the Eta Carinae binary system. 
The above dependence of the magnetic field strength on the distance from the star
has interesting consequences. The maximum energies of electrons, determined by the balance between their acceleration efficiency and the synchrotron energy losses, should increase proportionally with the distance of the shock from the stellar surface provided that the acceleration efficiency is independent ofthe magnetic field strength (see Eq.~8). Note however that at aertain distance from the stars the adiabatic losses can dominate over the hrotron energy losses.
On the other hand, the mum energies of electrons, determined by the IC energy losses in the T regime, should stay independent ofthe distance from the star (see Eq.~9). We conclude that if the acceleration process is saturated by the synchrotron energy losses, the
spectra of electrons should extend to higher energies for the parts of the shock farther from the star or for the apastron passage of stars when the whole shock structure is at a larger distance from the stars. 
However, the maximum energies should be independent of the shock location if the  electron acceleration is saturated by the IC energy losses in the T regime. This conclusion is valid provided that the acceleration efficiency of electrons does not depend on the distance from the stars and the advection time scale of electrons is always longer than the energy loss time scales.

\subsection{Hadrons}

On the other hand, hadrons lose energy mainly on collisions with the matter of the stellar winds. The time scale for the energy losses on pion production in proton-proton collisions can be estimated from 
\begin{eqnarray}
\tau_{\rm pp} = (\sigma_{\rm pp} k c \rho_{\rm w})^{-1}\approx 6.3\times 10^4R_{13}^2v_3/{\dot M}_{-4}~~~{\rm s},
\label{eq10}
\end{eqnarray}
\noindent
where $\sigma_{\rm pp}\approx 3\times 10^{-26}$ cm$^2$ is the cross section for
$p-p\rightarrow \pi$ production, $k = 0.5$ is the in-elasticity coefficient in this collision, and $\rho_{\rm w}$ is the density of the stellar wind at the shock region defined above. This energy loss time scale for the shock from the side of the Eta Carinae star is estimated to be $\sim 4\times 10^4$ s and from the side of the companion star as
$\sim 3.4\times 10^5$ s at the closest region of the shock to the stars and at their periastron passage. 
The time scales for energy losses by hadrons and their escape (advection) from the acceleration site do not depend on their energy. By comparing these time scales ($\tau_{\rm pp} = \tau_{\rm adv}$), we determine the condition under which hadrons can interact close to the shock within the binary system. The following condition has to be fulfilled
\begin{eqnarray}
{\dot M}_{-4} > 0.2R_{13}v_3^2.
\label{eq11}
\end{eqnarray}
This condition is clearly fulfilled at the periastron passage of the binary stars in the region of the shock from the side of the Eta Carinae star but is not fulfilled at the shock from the side of the companion star. Therefore, we expect that hadrons accelerated at the shock from the side of the Eta Carinae star interact efficiently within the binary system but those accelerated from the side of the companion star escape from the binary system to the surrounding nebula and interact there with the matter of expanding winds and the matter expelled during the past outbursts in the binary system (i.e., in the years 1843 and 1890).

The maximum energies of accelerated hadrons can be estimated by comparing the energy gains from the shock with energy losses on hadronic interactions or the advection from the acceleration site.
The comparison of $\tau_{\rm acc}$ with $\tau_{\rm pp}$ allows us to estimate the
maximum energies of hadrons
\begin{eqnarray}
E_{\rm p}\approx 6.3\xi_{-5}BR_{13}^2v_3/{\dot M}_{-4}~~~{\rm TeV}.
\label{eq12}
\end{eqnarray}
\noindent
If the condition given by Eq.~11 is not fulfilled, then the maximum energies of hadrons are determined by the escape along the shock. Then, the maximum energies are estimated as
\begin{eqnarray}
E_{\rm p}\approx 30\xi_{-5}BR_{13}/v_3~~~{\rm TeV}.
\label{eq13}
\end{eqnarray}

Hadrons are expected to be accelerated at the shock not only at the periastron passage but also at other phases of the binary system. The advection time of hadrons scales proportionally to the distance between the companion stars but the collision time of hadrons is only  proportional to the square of the distance (owing to the dependence of density of the wind with the distance from the star). We expect that hadrons are accelerated at the shock from the side of the Eta Carinae star to the maximum energies, which are independent on the phase of the binary system. However, their collision rate with the matter of the wind decreases at larger separation  of the stars within the binary system. On the other hand, hadrons, accelerated at the shock from the side the the companion star and farther from the periastron passage (or at the parts of the shock laying father from the star), should be accelerated to lower maximum energies. This is a consequence of the mainly radial structure of the magnetic field around the stars within the binary system ($B(r)\propto R^{-2}$). As a result, hadrons injected into the Eta Carinae nebula will have a
complicated spectrum, which can only be limited by the extreme values.
The more precise determination of this spectrum is difficult due to the unknown dependence of the efficiency of hadron acceleration in different parts of the shock and at different phases of the binary system.
Therefore, we calculate the expected radiation output below only from the hadrons at limiting distances from the star.

\section{Scenarios for gamma-ray production}

The complicated double shock structure of the collision region of the stellar winds, the possibile acceleration of electrons and hadrons, and the possible domination of different types of radiation mechanisms, that are responsible for the saturation of the acceleration process of particles provide the reach range of scenarios in which particles can produce high-energy radiation in the context of the massive binary systems.
Here we consider the production of high-energy radiation in some possible scenarios in the context of massive binary system with an elongated orbit.
They will be tested by the available high-energy observations of Eta Carinae.
They also predict some specific features, which will allow us to distinguish between them with the future more systematic X-ray and GeV-TeV $\gamma$-ray and neutrino observations. 

At first, we conclude that the hard X-ray emission detected by Beppo-SAX, INTEGRAL, and SUZAKU cannot be produced by electrons in the synchrotron process. The maximum energies of synchrotron photons produced by electrons whose acceleration is saturated by the synchrotron energy losses can be estimated from
\begin{eqnarray}
\varepsilon \approx m_{\rm e}(B/B_{\rm cr})\gamma_{\rm e}^2\approx 1.3\xi_{-5}
~~~{\rm keV},
\label{eq14}
\end{eqnarray}
\noindent
where $\gamma_{\rm e}$ is the Lorentz factor of electrons estimated from Eq.~8, and $B_{\rm cr} = 4.4\times 10^{13}$ G is the critical magnetic field. Because the maximum energies of observed hard X-ray emission is at least $\varepsilon\sim 0.1$ MeV, the acceleration efficiency has to fulfil the condition $\xi_{-5} > 76$ (see Eq.~\ref{eq14}). This condition cannot be fulfilled in the shocks either from the side from the companion star or from the side of the Eta Carinae star. If the acceleration of primary electrons is saturated by the synchrotron energy losses, we should observe a strong synchrotron component in the Eta Carinae spectrum from the central source peaking at $\sim 1$ keV. This is in-consistent with the observations of Eta Carinae because the detected soft X-ray emission has a thermal nature and is clearly extended.
Therefore, we conclude that the hard X-ray emission cannot be produced by electrons in the synchrotron process, which are directly accelerated at the shocks. 

Below we consider the two most likely scenarios for the high-energy emission from the Eta Carinae binary system, i.e., the production of the whole $\gamma$-ray spectrum only by electrons (our model A) and the production of the $\gamma$-ray spectrum in a composite model, part of the spectrum by electrons and part by hadrons (model B).

\subsection{Model A: electrons}

The non-thermal multiwavelength spectrum measured from the Eta Carinae binary system has its maximum in the $\gamma$-ray energy range. We expect that the energy losses of accelerated electrons in the synchrotron process are less than their energy losses in the IC process. As we noted above,
both shocks present within the binary system are characterised by different values of the acceleration efficiency ($\xi_{\rm comp}\sim 10^{-4}$ and $\xi_{\rm EC}\sim 5\times 10^{-6}$) and different values of the magnetic field strength at the shock.
We fixed the values for the surface magnetic field of the companion star as 
$B_{\rm comp} = 2\times 10^3$ G and of the Eta Carinae star as $B_{\rm EC} = 200$ G, and assumed a mainly radial structure of the magnetic field in the stellar winds within the binary system (i.e. $B(r)\propto r^{-2}$). With these assumptions, we estimated the magnetic field strength at the shock from the companion star to be $B\approx 60$ G, and from the Eta Carinae star as $B\approx 100$ G at the periastron passage. 
We compare the characteristic time scales important for acceleration of electrons and their energy losses for the parameters mentioned above (see Fig.~2). The  acceleration of those electrons at the parts of the shock that are the closest to the companion stars at their periastron passage is considered.
We find that electrons can be accelerated at the
shock from the side of the companion star to the maximum energies $E_{\rm e}^{\rm max}\approx 80$ GeV. These energies are owing to the saturation by the synchrotron energy losses (Fig.~2). 
The saturation of electron acceleration by the synchrotron energy losses happens for energies close to the transition between the dominance of the synchrotron and the IC energy losses in the KN regime. Therefore, in spite of the saturation of electron acceleration by synchrotron process, these electrons lose energy mainly on the IC scattering in the KN regime (see Fig.~2).
The acceleration process of electrons at the parts of the shock at larger distances from the stars should also be saturated by the synchrotron energy losses because their acceleration time scale depends on the distance from the stars in a similar way as their energy loss time in the KN regime (i.e. $\tau_{\rm acc}\propto \tau_{\rm IC/KN}\propto R^2$, see Eqs.~2 and~6). However, the maximum energies of electrons accelerated at the parts of the shock at larger distances from the stars should increase proportionally to R (see Eq.~8).

The maximum energies of electrons at the shock from the side of the Eta Carinae star are determined by the balance between the acceleration process and the IC energy loss process in the T regime. For the parameters mentioned above, they are estimated to be $E_{\rm e}^{\rm max}\approx 7$ GeV (see Eq.~9). These maximum energies should not change in the parts of the shock farther from the stars or at other phases because they are independent of the distance from the stars (see Eq.~9).
Electrons accelerated at the shock from the side of the Eta Carinae star lose energy mainly on the IC scattering of stellar radiation in the Thomson regime. 
The energy density of the radiation at the shock region from the Eta Carinae star and the companion star are comparable. Therefore, $\gamma$-rays produced in the IC process should have a quite isotropic distribution.

In conclusion, we expect the presence of two components in the electron spectrum, which are accelerated by different shocks created in the collision of the winds from the sides of both stars. These electrons can cool efficiently in the radiation field of the companion stars. These two components in the electron spectrum are responsible in our Model A for the two-component $\gamma$-ray spectra observed from the Eta Carinae binary system.

\begin{figure}
\vskip 7.truecm
\includegraphics{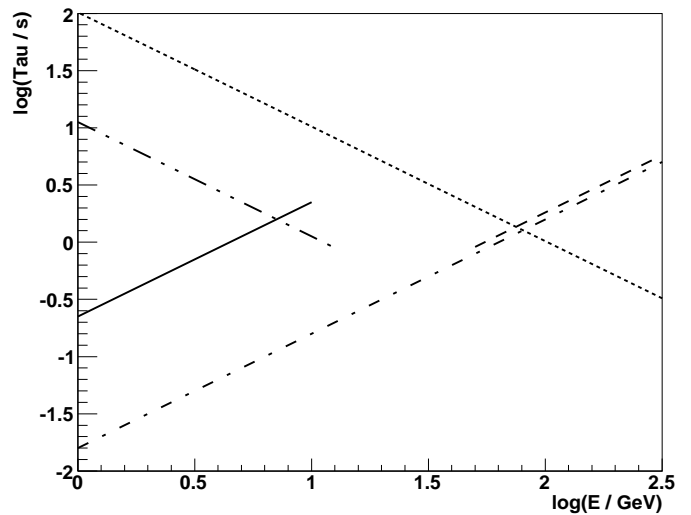}
\caption{Characteristic time scales for electrons in the acceleration process
in the shock from the side of the companion star (dot-dashed line) and the side of the Eta Carinae star (solid), the synchrotron cooling time of electrons at the shock from the side of the companion star (dotted), the IC cooling time scale in the Thomson regime (dot-dot-dashed) and in the Klein-Nishina regime (dashed). The intersection of the acceleration time scale with the IC cooling time scale in the Thomson regime gives the maximum energies of electrons accelerated at the shock from the side of the Eta Carinae star. The intersection of the acceleration time scale  with the synchrotron cooling time scale gives the maximum energies of electrons accelerated at the shock from the side of the companion star. The parameters defining the acceleration scenario are $B_{\rm sh}^{\rm comp} = 60$ G, $B_{\rm sh}^{\rm EC} = 100$ G, $\xi^{\rm comp} = 10^{-4}$, and  $\xi^{\rm EC} = 5\times 10^{-6}$.}
\label{fig2}
\end{figure}
\subsection{Model B: electrons and hadrons}

In the second model we assume that the sub-GeV $\gamma$-rays are produced by electrons as in the case of model A. However, the second higher energy component in the $\gamma$-ray spectrum is expected to be produced by hadrons (see also Leyder et al.~2010).
Acceleration of hadrons at the shock from the side of the Eta Carinae star is saturated by their energy losses on pion production. Their maximum energies are estimated from Eq.~(12) on $E_{\rm p}^{\rm max}\approx 250$ TeV. These hadrons cool close to the acceleration site within the binary system, producing high-energy $\gamma$-rays and neutrinos.
On the other hand, acceleration of hadrons at the shock from the side of the companion star is limited by their convection along the shock. Then, the maximum energies of hadrons in the part of the shock that is the closest to the stars at the periastron passage, are estimated to be $E_{\rm p}^{\rm max}\approx 5$ PeV (see Eq.~13). These hadrons escape to the nebula surrounding the Eta Carinae binary system and eventually interact with the matter within the nebula, producing high-energy gamma-rays and neutrinos.
In principle, hadrons accelerated in both shocks can contribute to the observed hard $\gamma$-ray emission. However, their relative contribution
is difficult to estimate because of it depends on the details of the acceleration process, which are mainly unknown.  On the other hand, hadrons accelerated at the shock from the side of the companion star accumulate within the nebula. They
interact with the matter of expending winds of the stars and the matter entrained within the nebula during the past outbursts.

\begin{figure*}
\vskip 6.4truecm
\includegraphics{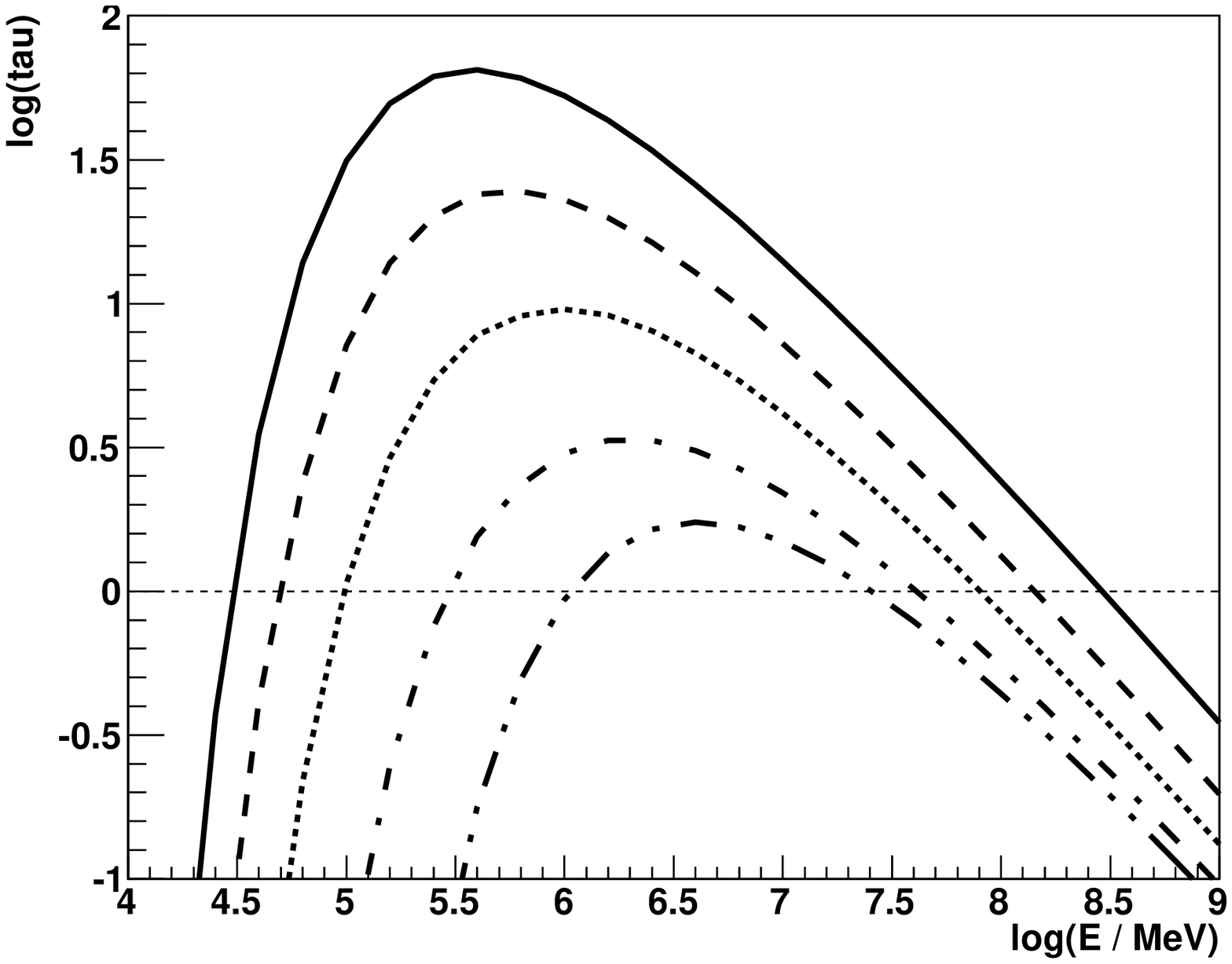}
\includegraphics{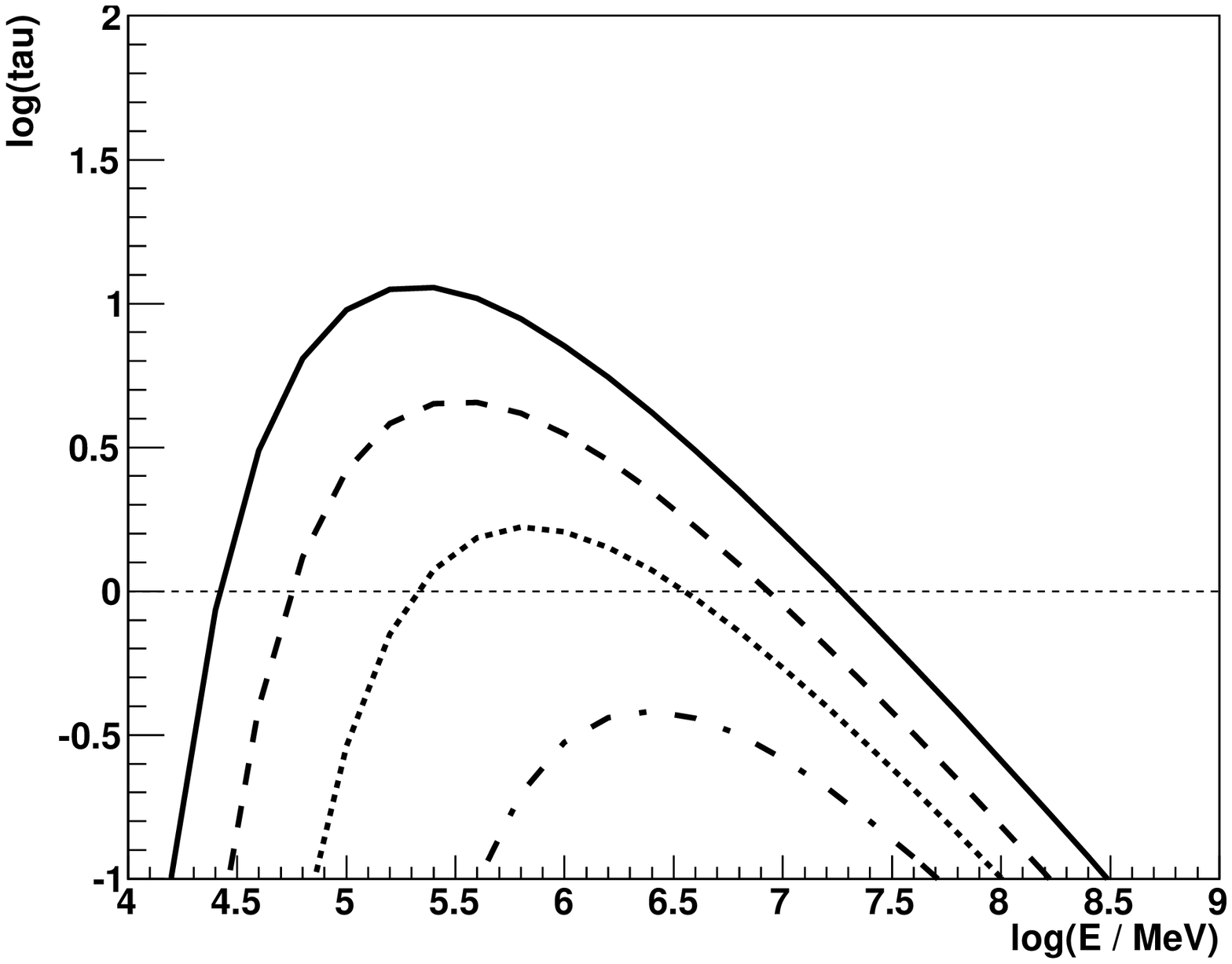}
\includegraphics{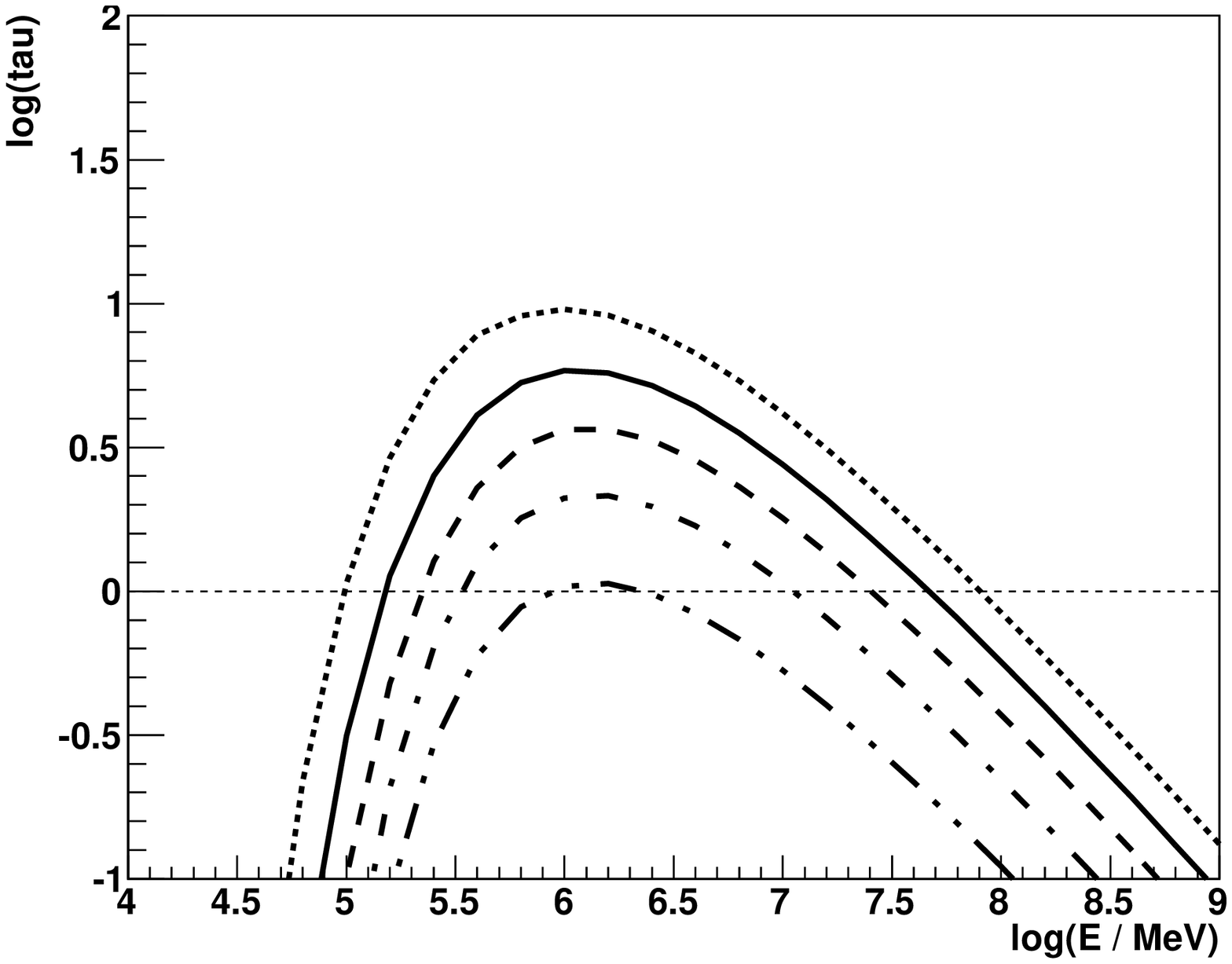}
\caption{Optical depths for $\gamma$-rays in the radiation field of the Eta Carinae star (left panel) and its companion star (middle panel) as a function of energy of injected $\gamma$-ray photons. Specific curves show the results for the propagation angle of the $\gamma$-ray photon (measured from the direction defined by the injection place and the center of specific star) equal to: $\alpha = 0^{\rm o}$ (dot-dot-dashed curve, direction outwards of the specific star), $30^{\rm o}$ (dot-dashed), $60^{\rm o}$ (dotted), $90^{\rm o}$ (dashed), $120^{\rm o}$ (solid). $\gamma$-rays are injected at the distance of $1.4R_{EC}$ from the surface of Eta Carinae and at the distance of $6R_{\rm comp}$ from the surface of the companion star. The dependence of the optical depths on the distance from the Eta Carinae star is shown in the right panel for the injection angle of $\alpha = 60^{\rm o}$ and the distances $1.4R_{EC}$ (dotted), $2R_{EC}$ (solid), $3R_{EC}$ (dashed), $5R_{EC}$ (dot-dashed), and $10R_{EC}$ (dot-dot-dashed), respectively.}
\label{fig3}
\end{figure*}

Note that in principle the efficiencies of the particle acceleration in both shocks can differ as can the efficiencies of the acceleration of electrons and hadrons. Therefore, a different combination of processes can produce a variety of radiation outputs from these binary systems.

\section{Absorption of gamma-rays}

The TeV $\gamma$-rays produced within the compact binary systems should be absorbed in a complicated way in the soft radiation of luminous stars (e.g. Bednarek~1997).
These absorption effects should be taken into account for the case of the $\gamma$-ray production within $\sim$10 stellar radii from both stars. 
Therefore, we calculated the optical depths for $\gamma$-rays injected close to the shock region in the radiation field of both stars as a function of their propagation angle and for their different injection distances from the stars.
We applied the parameters of both stars as mentioned in the introduction.
The optical depths are shown in Fig.~\ref{fig3}. The angles for the Eta Carinae star and the companion star are counted outwards from the specific star. 
Clearly even in the most favourable case defined by the angle $\alpha = 90^o$, the $\gamma$-rays with energies above a few tens of GeV should be efficiently absorbed (the optical depth above unity). 
These absorption effects should be taken into account when comparing the $\gamma$-ray spectra produced close to the periastron passage of the stars with the observations of the Eta Carinae binary system.
However, closer to the apastron passage, most of the shock structure lies at distances larger than $\sim 10$ stellar radii. At these distances the optical depths drop below unity (see Fig.~3 for the dependence of the optical depth on the distance).
Therefore, it is expected that absorption of $\gamma$-rays produced close to the shock region will result in a clear modulation of the $\gamma$-ray signal from the Eta Carinae binary system already at sub-TeV energies. 

\section{Comparison with observations of Eta Carinae}

We discuss the radiation output expected in terms of the model A and B (defined above) for different locations of the acceleration site from the companion stars, i.e. for different parts of the shock and different phases of the binary system. The calculated hard X-ray to $\gamma$-ray spectra are compared with the recent observations of the Eta Carinae binary system (described in the introduction).

\subsection{Model A}

In this model we assumed that the GeV $\gamma$-ray emission is produced by electrons accelerated at the shock from the side of the Eta Carinae star and the multi-GeV hard $\gamma$-ray emission is produced by electrons accelerated at the shock from the side of the companion star. As an example, we assume that electrons are accelerated with the power law spectrum defined by the spectral index $\alpha_{\rm inj} = 2$ up to the maximum energies as estimated in Sect.~3.1.

For electrons accelerated at the shock from the side of the Eta Carinae star, the main energy loss mechanism
is the IC scattering in the T regime. This process is efficient enough already close to the acceleration site because the cooling time scale of electrons on the IC process is clearly shorter than their advection time scale
(see Eqs.~3 and 5). As a result, the equilibrium spectrum of electrons reaches the spectral index $\alpha_{\rm eq} = 3$ and the $\gamma$-ray spectra from these electrons should have a spectral index close to $2$. However, at low electron energies, the cooling time scale of the IC process starts to become comparable to the cooling time scale of the bremsstrahlung process (see Eqs.~5 and 7). The transition between which of these two radiation processes dominates occurs for electrons with energies of a few tens of MeV. Therefore, the photon spectrum from the IC process should steepen at energies somewhere in the soft $\gamma$-ray to hard X-ray energy range. The precise break energy in the IC spectrum is difficult to estimate because of the uncertain density of the matter in the shock region (it varies by a factor of $\sim$100, see  Pittard~2009). For low-energy electrons, for which the  bremsstrahlung energy losses dominate, the equilibrium spectrum
should be close to the injection spectrum $\alpha_{\rm eq}^{\rm brem} = \alpha_{\rm inj}$. Therefore, the IC spectrum produced by these electrons should steepen to the spectral index $\sim$1.5.
We compare the hard X-ray to $\gamma$-ray IC spectrum expected from the electrons accelerated at the shock from the side of the Eta Carinae star with the observations of this binary system in Fig.~4 (see solid curve in the left panel). $\gamma$-rays from the bremsstrahlung process are produced with energies comparable to energies of their parent electrons. The power in this emission only contains a small part of the whole power of accelerated electrons. Therefore, its contribution to the spectrum observed from Eta Carinae is on a much lower level.

\begin{figure*}
\vskip 6.5truecm
\includegraphics{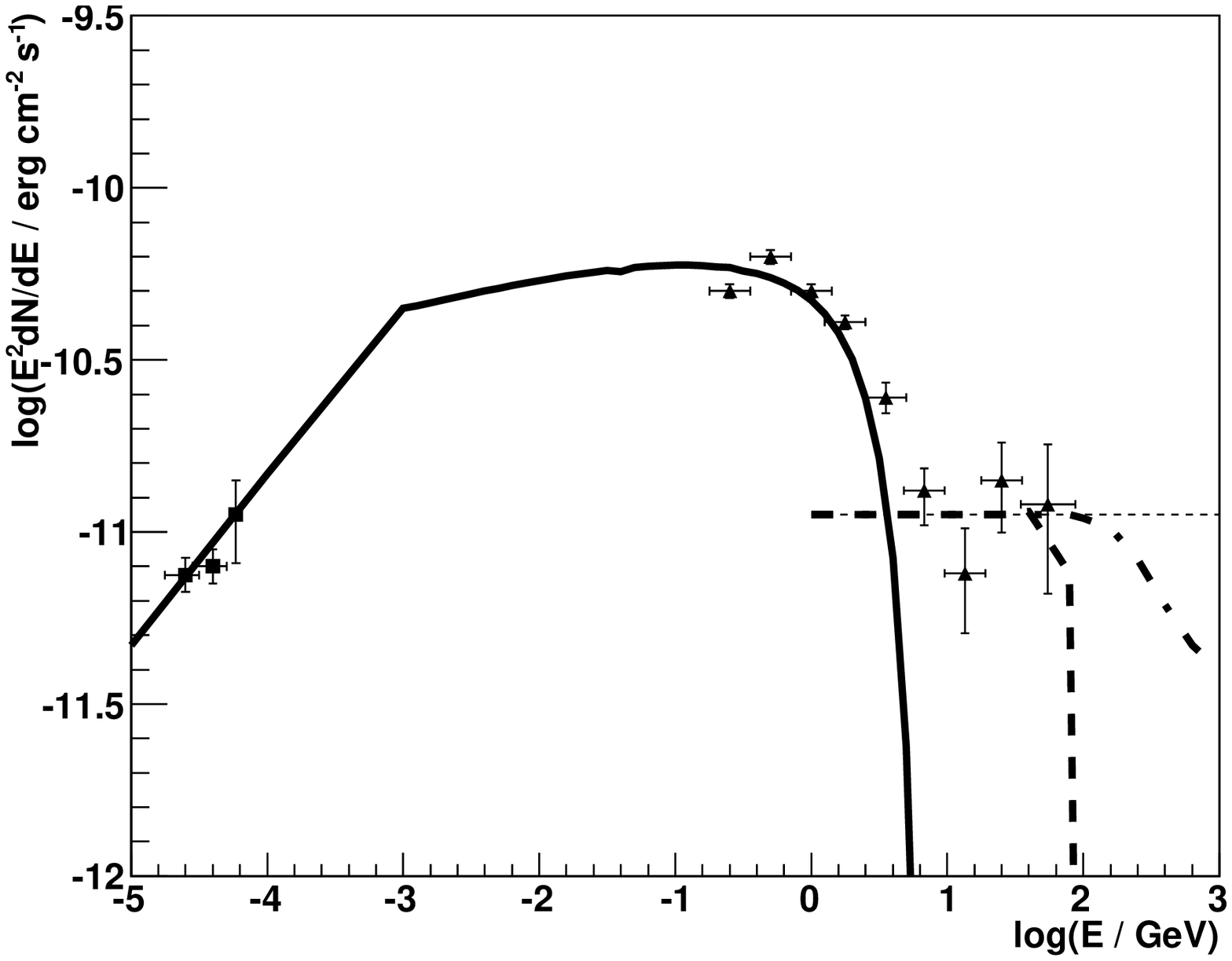}
\includegraphics{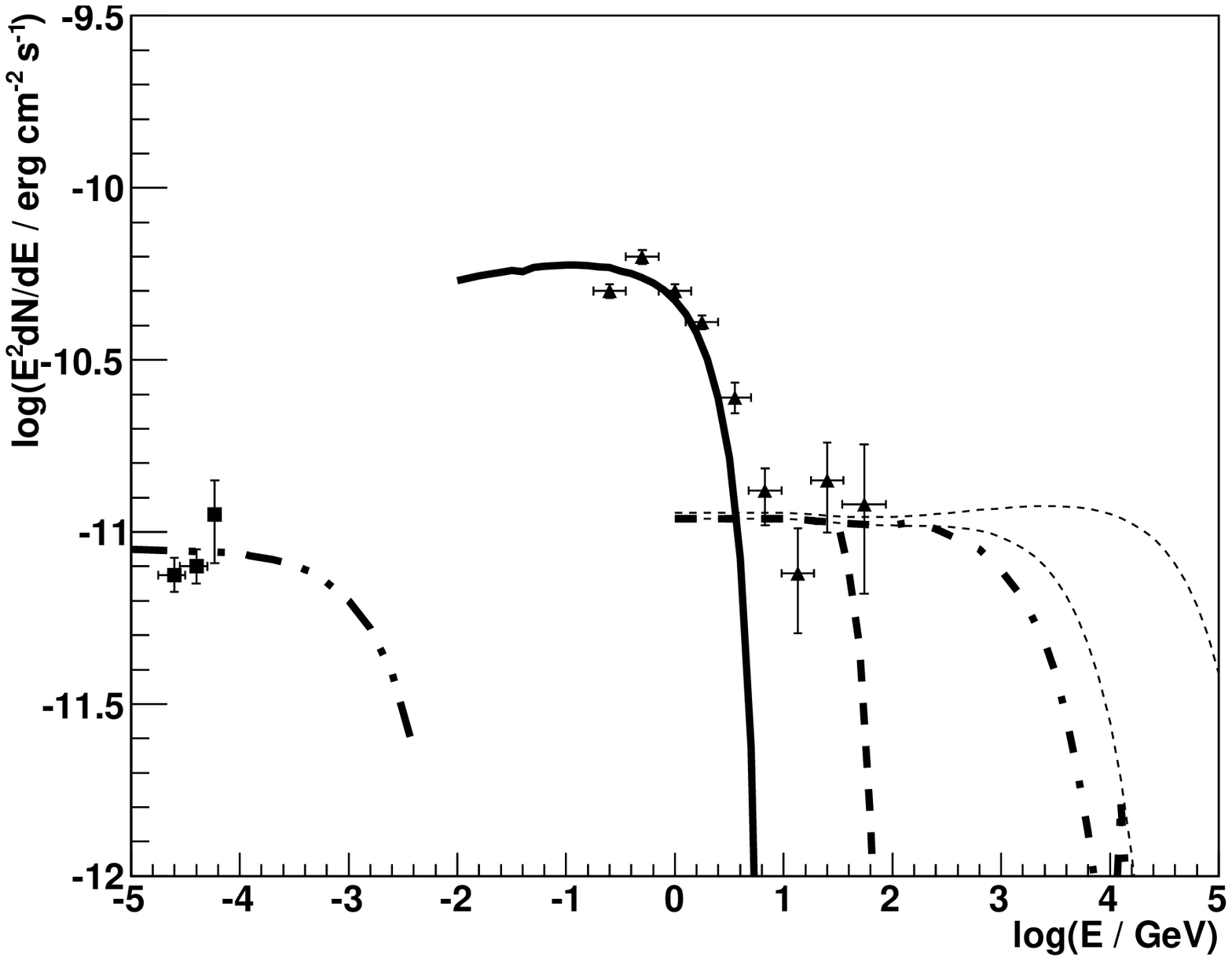}
\caption{X-ray and $\gamma$-ray spectra expected for the model A (left) and model B (right) at the periastron passage of the stars in the binary system Eta Carinae  compared with the high-energy observations of this system. The measurements in the hard X-rays by the INTEGRAL satellite (Leyder et al.~2010) are marked by squares and those in the $\gamma$-ray energy range by the Fermi-LAT detector (Walter et al.~2010, Abdo et al.~2010) are marked by triangles. The results of the calculations are shown for the model A (left panel): $\gamma$-rays from electrons accelerated at the shock from the side of the Eta Carinae star (solid) and from electrons accelerated at the shock from the side of the companion star: without any absorption (thin dashed), with absorption for the $\gamma$-ray propagating at the angle equal to $50^{\rm o}$ (towards the star) at distance from the star $D=1.4R_{\rm EC}$ (dashed),
and at the same angle but for the parts of the shock at the distance from the star $D = 10R_{\rm EC}$ (dot-dashed). For Model B (right panel): $\gamma$-rays from electrons, accelerated at the shock from the side of the Eta Carinae star, scattering stellar radiation in the T regime (solid), $\gamma$-rays from decay of neutral pions produced by hadrons accelerated at the shock from the side of the Eta Carinae star: unabsorbed spectrum  (thin dashed curve extending to lower energies), spectra absorbed in the stellar radiation for distances from the star: $D = 1.4R_{\rm EC}$ (thick dashed) and $D = 10R_{\rm EC}$ (dot-dashed). Secondary leptons from decay of charged pions (produced in hadronic collisions) lose energy on the synchrotron radiation 
(dot-dot-dashed curve). The $\gamma$-ray spectrum produced by hadrons, which are accelerated at the shock from the side of the companion star and produce pions in the interactions within the Eta Carinae nebula, are shown by the thin dashed curve extending to higher energies.}
\label{fig4}
\end{figure*}

On the other hand, electrons accelerated at the shock from the side of the companion star can reach energies of the order of $\sim 100$ GeV. The electron energies  can be even higher, provided that their acceleration occurs in the part of the shock at larger distances from the stars (Sect.~3.1). Electrons also lose energy in the IC scattering process, which in this case occurs in the T and in the KN regimes (the synchrotron energy losses are neglected, see Fig.~2 for details of the derivation of maximum energies of accelerated electrons).
We show that this IC emission can be responsible for the the hard multi-GeV $\gamma$-ray spectrum observed from the direction of the Eta Carinae binary system. 
For the injection spectrum of electrons with the spectral index $\alpha_{\rm inj} = 2$, the IC $\gamma$-ray spectrum is also produced with the spectral index
$\alpha_\gamma = 2$, both in the T and KN regimes (see e.g. Blumenthal \& Gould~1970). However, as we showed above, $\gamma$-rays produced at the part of the shock relatively close to the companion stars are efficiently absorbed in the stellar radiation field. We take these absorption effects into account by applying the optical depths for the $\gamma$-rays in the radiation fields of the Eta Carinae and companion stars
and using simple absorption law, $\propto \exp[-\tau(E_\gamma)]$. Because the inclination angle of the Eta Carinae binary system is estimated to be $\sim 40^{\rm o}$, we show as an example the $\gamma$-ray spectra with the absorption effects in the radiation of the Eta Carinae star for the angle $\alpha = 50^{\rm o}$ (see dashed and dot-dashed curves in left Fig.~4).
The unabsorbed $\gamma$-ray spectrum (produced at distances $>$10$R_{\rm EC}$ from the Eta Carinae star) are shown by the dotted curves. 
We conclude that the considered model predicts a clear modulation of the $\gamma$-ray signal at energies above $\sim$100 GeV with the period of the binary system. The $\gamma$-ray spectrum produced at the periastron passage by electrons accelerated at the part of the shock that are the closest to the Eta Carinae star should cutoff just below $\sim 100$ GeV. 

\subsection{Model B}

In this model the GeV $\gamma$-ray peak is also produced by electrons
accelerated at the shock from the side of the Eta Carinae star (as in model A).
But the multi-GeV hard $\gamma$-ray emission is proposed to be produced by hadrons that are also accelerated at the shock from the side of the Eta Carinae star (see also Walter et al.~2010). At the periastron passage the hadrons accelerated at this shock can reach energies $\sim 250$ TeV for the magnetic field strength at the shock from the side of the Eta Carinae star (see Eq.~12). These maximum energies are independent of the distance from the stars in the case of the radial structure of the magnetic field.
The number of collisions of relativistic hadrons with the matter of the Eta Carinae wind, already close to the acceleration site within the binary system, can be estimated as, 
\begin{eqnarray}
N_{\rm bin}^{\rm col}\approx \tau_{\rm adv}/\tau_{\rm pp}\approx 
4.8{\dot M}_{-4}/(v_3^2R_{13}).
\label{eq15}
\end{eqnarray}
\noindent
For the parts of the shock close to the Eta Carinae star at the periastron passage, the collision rate is estimated to be $\sim$15, and at the distance of $10R_{\rm EC}$ it is still $\sim$1.5.
Therefore, we conclude that these hadrons are efficiently cooled already close to the shock within the binary system through the large range of the binary phases. As a consequence, $\gamma$-ray spectra from the decay of the neutral pions produced by hadrons accelerated at the Eta Carinae shock should not depend strongly on the phase of the binary system. We calculate the spectra of $\gamma$-rays and secondary leptons from the decay of the pions produced in hadronic collisions assuming the power law spectrum of hadrons with the spectral index
$\alpha_{\rm p} = 2$. In Fig.~4, $\gamma$-ray spectra from decay of $\pi^o$ are compared with the $\gamma$-ray observations of the Eta Carinae binary system. 

The $\gamma$-ray spectrum from decay of pions clearly extends through the TeV energy range. However, as in the previous model, the TeV $\gamma$-rays produced within $\sim$10 stellar radii should be efficiently absorbed in the thermal radiation of the Eta Carinae star (Sect.~5). We take this absorption effect into account by showing in Fig.~4 the unabsorbed $\gamma$-ray spectra (thin dashed curves) and absorbed spectra for the parameters of the binary system as for model A. The maximum energies of the accelerated hadrons should not depend on the distance from the stars for the radial structure of the magnetic field, provided that the acceleration coefficient does not depend on the distance (see Eq.~12). Therefore, at locations close to apastron, the $\gamma$-ray emission extending up to a few TeV is in principle possible. 
We conclude that the Cherenkov telescopes should detect a clear modulation of the $\gamma$-ray signal at energies above $\sim$100 GeV with the orbital period of the binary system because of the different absorption conditions within the binary system.

The secondary leptons, from decay of charged pions produced in hadronic collisions, efficiently lose energy in the synchrotron process, reaching the equilibrium spectrum with the spectral index $\alpha_{\rm e^\pm}^{\rm eq} = 3$. They produce synchrotron radiation with the spectral index $\alpha_{\rm syn}^{\rm e^\pm} = 2$. The spectral index of the hard X-ray emission from the Eta Carinae binary system is not precisely known (Viotti et al.~2004, Leyder et al.~2008, Leyder et al.~2010, Sekiguchi et al.~2009). However, it is reported to be not far from the above value.
The synchrotron emission from secondary leptons should extend through the hard X-ray energy range. The spectrum of secondary leptons, from decay of charged pions produced in hadronic collisions, extends up to
\begin{eqnarray}
E_{e^\pm}^{\rm max}\approx E_{\rm p}^{\rm max}/(8\mu)\sim 1~~~{\rm TeV}, 
\label{eq16}
\end{eqnarray}
\noindent
where $\mu$ is the multiplicity of pion production by protons given in Orth \& Buffington~(1976). Electrons with these energies ($E_{\rm e} = m_{\rm e}\gamma_{\rm e}$) are able to produce synchrotron radiation at the shock from the side of the Eta Carinae star with energies up to
\begin{eqnarray}
\varepsilon_{\rm syn} \approx m_{\rm e}(B/B_{\rm cr})\gamma_{\rm e}^2\sim 5
~~~{\rm MeV},
\label{eq17}
\end{eqnarray}
i.e., consistent with the observations of the hard X-ray emission from the Eta Carinae binary system (see Fig.~4). The level of this emission should also be comparable to the level of the hard $\gamma$-ray component from Eta Carinae, which is indeed the case (see Abdo et al.~2010 and Walter et. al.~2010). Therefore, secondary leptons can contribute to (or even be responsible for) the hard X-ray emission from the Eta Carinae binary system.
The spectral index of the IC component in the hard X-ray energy range produced by primary electrons accelerated at the shock from the side of the Eta Carinae star is predicted to be clearly flatter than 2 (see Model A). Therefore, precise measurements in this energy range should indicate the origin of this lower energy emission, i.e., either it is produced by primary electrons in the IC process or by secondary leptons in the synchrotron process.

In principle, the hard $\gamma$-ray component could be also produced by hadrons accelerated at the shock from the side of the companion star. However, these hadrons escape to the Eta Carinae nebula from the binary system without significant energy losses close to the acceleration site. The Larmor radii of hadrons escaping to the nebula are always much smaller than the characteristic distance scale of the system (which is of the order of the distance from the binary system) for the magnetic field around the massive stars with the toroidal structure at a distance above $\sim 10R_{\rm EC}$ (USO \& Melrose~1992). Hadrons are frozen in the stellar winds moving outwards from the binary system.
We can distinguish two regions within the nebula in which hadrons interact with the matter, i.e., the inner region where the density of matter is dominated by the direct stellar winds and the outer region where the density of the wind drops below the average density of the Eta Carinae nebula created during the outburst about 170 years ago. 
We assume that at larger distances from the binary system the winds from the Eta Carinae star and the companion star can mix through  the rotation of the binary system. The characteristic collision rate of hadrons can be estimated from
\begin{eqnarray}
N_{\rm tot}^{\rm col} =
\int_{R_{\rm bin}}^{R_{\rm neb}}{{\rho_{\rm w}(r)\sigma_{\rm pp}c}\over{v_{\rm w}}}dr + 
\tau_{\rm neb}\rho_{\rm neb}\sigma_{\rm pp}c.
\label{eq18}
\end{eqnarray}
\noindent
where $R_{\rm bin}$ is the radius of the binary system assumed to be equal to $\sim 10^{14}$ cm, $R_{\rm neb} = 3\times 10^{17}$ cm, $\rho_{\rm w}(r)$ is the density of the wind from the binary system (see above), $v_{\rm w}$ is the velocity of the stellar wind, $\rho_{\rm neb}$ is the average density of the nebula $\sim 10^4$ cm$^{-3}$, and $\tau_{\rm neb}\approx 5\times 10^9$ s is the age of the Eta Carinae nebula. For the applied parameters, $N_{\rm tot}^{\rm col}$ can reach the values of a few, i.e., hadrons can completely cool within the nebula. 
In principle, hadrons injected into the expanding wind from the star could suffer from adiabatic energy losses. To check whether this process can significantly change the energies of the injected hadrons we estimate the distance, $R_{\rm out}$, from the binary system up to which hadrons interact at least once with the matter of the wind. The following equation has to be checked: $\int_{R_{\rm bin}}^{R_{\rm out}}{{\rho_{\rm w}(r)\sigma_{\rm pp}c}\over{v_{\rm w}}}dr = 1$. For the example parameters applied in our calculations we estimate that hadrons should mainly interact relatively close to the binary system, where the density of the wind is still high. Therefore, adiabatic energy losses of hadrons through the expansion of the wind can be safely neglected.
We conclude that hadrons injected into the nebula surrounding the Eta Carinae binary system should interact efficiently, producing $\gamma$-rays and neutrinos. This conclusion is in contrast to the expectations of Ohm et al.~(2010), who predicted inefficient interaction of hadrons in the large-scale nebula with relatively low average density. However,  
we consider the inner, denser nebula with the radius of a few times smaller than applied in their case. This inner nebula is dominated by the matter from the dense wind of the Eta Carinae star produced at the present time.

The maximum energies of hadrons escaping from the binary system can be as high as
$\sim$5 PeV (see Eq.~13) for the part of the shock that is closest to the companion star at the periastron passage. These maximum energies drop with the distance of the shock from the companion star if the radial
dependence of its magnetic field reaches values of $\sim$250 TeV for the distance of $20R_{\rm EC}$. As an example, we calculate the $\gamma$-ray spectrum expected from hadrons within the nebula with the power law spectrum, applying the  spectral index equal to $2$ and the maximum energies of hadrons equal to 5 PeV (Fig.~4). If hadrons are accelerated efficiently at the shock from the side of the companion star, a $\gamma$-ray emission extending up to multi-TeV energies is expected. This emission component should be steady, i.e., independent of the phase of the binary system because the absorption effects in the stellar radiation fields within the Eta Carinae nebula are not important. 

The total energy content in relativistic hadrons, which are responsible for the observed $\gamma$-ray luminosity, can be estimated from,
$P_{\rm tot}^{\rm p}\approx 3L_\gamma^{\rm EC} \tau_{\rm pp}$, where $L_\gamma^{\rm EC}$ is the $\gamma$-ray luminosity observed from the Eta Carinae binary system above $\sim$10 GeV, $\tau_{\rm pp}$ is the cooling time scale for hadronic collisions given by Eq.~(10), and the factor of 3 is there because only about one third of the produced pion energy goes to $\gamma$-rays.   
For $L_\gamma^{\rm EC}\approx 2\times 10^{34}$ erg s$^{-1}$ derived for the distance to Eta Carinae equal to 2 kpc (Abdo et al.~2010) and the density of matter characteristic at the border of the binary system, $\rho_{\rm w}(R_{\rm bin} = 10^{14} cm)\sim 10^9$ cm$^{-3}$, we estimate the total energy content in relativistic hadrons within the inner nebula to be $P_{\rm tot}^{\rm p}\sim$10$^{41}$ erg. Because the energy losses of hadrons are very efficient in this model, hadrons reach a steady-state distribution in the inner nebula. Then, the power radiated in $\gamma$-rays is comparable to the power converted from the acceleration mechanism to relativistic hadrons, $L_{\rm p}\approx L_\gamma^{\rm EC}$. We conclude that to provide the hard $\gamma$-ray luminosity observed from the Eta Carinae binary system, hadrons have to take only 
$\sim$10$^{-3}$ of the power of the wind from the Eta Carinae star, which is estimated to be $L_{\rm w} = {\dot M}v_{\rm w}^2/2\approx 2\times 10^{37}$ erg s$^{-1}$.

Hadrons might also be accelerated at the large-scale blast wave produced during the past explosion of the Eta Carinae star (e.g. Ohm, Hinton \& Domainko~2010). This injection process can provide an additional source of high-energy hadrons within the nebula which was not considered in the present paper.

\section{Neutrinos from Eta Carinae}

Hadrons accelerated at both shocks should also produce neutrinos in collisions with dense matter of the Eta Carinae wind and the matter entrained within the Eta Carinae nebula during past outbursts.
In model B, we propose that hadrons accelerated at the shock from the side of the Eta Carinae star can interact efficiently with the matter of the stellar wind, producing $\gamma$-rays via decay of pions from hadronic interactions through the large range of the binary phases. The maximum energies of accelerated hadrons are estimated to be $\sim 250$ TeV. They should not depend strongly on the distance of the shock from the Eta Carinae star. We calculate the muon neutrino spectra produced by these hadrons by applying the scaling break model, which is suitable for hadrons with the above estimated energies (Wdowczyk \& Wolfendale~1987). It is assumed that hadrons have a power law spectrum with the spectral index equal to $2$. Such a flat spectrum is motivated by the very flat hard GeV $\gamma$-ray emission observed from the Eta Carinae. We compare the neutrino spectra with the atmospheric neutrino background and the sensitivity of 1 km$^2$ neutrino detector of the IceCube type (see Fig.~5). Unfortunately, the level of this neutrino emission is comparable to the level of the atmospheric neutrino background. Therefore, these neutrinos will be difficult to detect even with a 1 km$^2$ neutrino detector.

\begin{figure}
\vskip 6.4truecm
\includegraphics{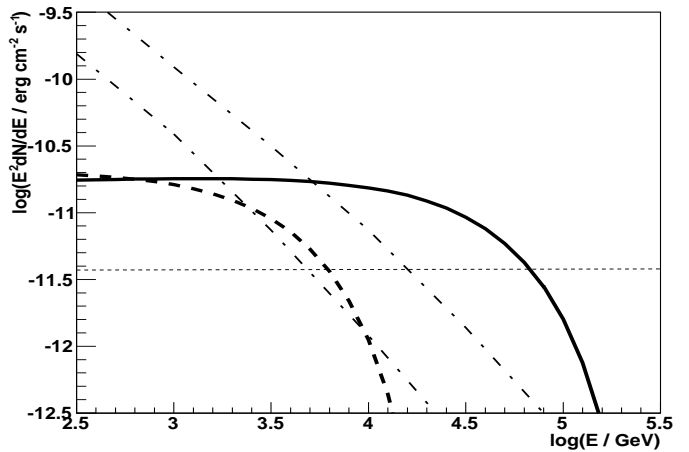}
\caption{Muon neutrino spectra produced by hadrons which are accelerated at the shocks from the side of the companion star up to energies of 5 PeV (solid curve) and the side of the Eta Carinae star up to energies of 250 TeV (dashed). Hadrons from the side of the Eta Carinae star interact with the matter of dense wind close to the shock region within the binary system. Hadrons from the side of the companion star escape from the shock region into the nebula and interact with the matter of expanding stellar wind and the matter within the nebula expelled in the past. The power law spectrum of hadrons is applied with the spectral index equal to $2$. The neutrino spectra are normalized to the hard $\gamma$-ray spectrum observed from the Eta Carinae binary system assuming its hadronic origin via pion production. 
The range of the atmospheric neutrino background (vertical and horizontal) within $1^{\rm o}$ degree of the source is marked by dot-dashed curves (Lipari~1993). The 1 yr sensitivity of the km$^2$ neutrino detector (expected for the IceCube detector) is marked by the thin dotted line.}
\label{fig5}
\end{figure}

Hadrons accelerated at the shock from the side of the companion star can also produce neutrinos in collisions with the matter of the Eta Carinae nebula.
We show that this scenario can be tested by the future neutrino observations of the Eta Carinae binary system. By assuming that $\gamma$-rays, also  produced by these hadrons, are responsible for the observed hard GeV $\gamma$-ray emission from the Eta Carinae binary system, we estimate the expected muon neutrino spectrum from hadrons accelerated at the shock from the side of the companion star. The neutrino signal should result as a contribution from hadrons accelerated at the shock during the whole range of phases of the binary system. Their maximum energies at these  different locations vary between 5 PeV (periastron) to 250 TeV (at distance of the 
shock equal to $D = 20R_{\rm EC}$). Therefore, the total neutrino flux from hadrons, accelerated at the shock from the side of the companion star and interacting with the matter of the Eta Carinae nebula, should be contained within the range    
defined by the dashed and solid curves in Fig.~5. There is a chance that
these neutrinos would be detected by the 1 km$^2$ detector because the predicted spectrum is above the atmospheric neutrino background and above the sensitivity of the 1 km$^2$ neutrino detector at energies between a few TeV and a few tens of TeV. Non-detection of the muon neutrino signal in this energy range (or detection on lower level) will give us important information on the acceleration efficiency of hadrons at the shock from the companion star.

\section{Discussion and conclusion}

We investigated the consequences of the acceleration of particles (electrons, hadrons) at the shocks appearing within the binary system of two massive stars as a result of stellar wind collisions. 
In contrast to previous studies, we considered the acceleration of particles in two shock structures which appear on both sides of the contact discontinuity from the sides of the Eta Carinae star and the companion star. These two shocks have different properties owing to differences in the surface magnetic field strengths
($B_{\rm EC} = 200$ G and $B_{\rm comp} = 2\times 10^3$ G) and acceleration efficiencies depending on the stellar wind velocities ($\xi_{\rm EC} = 5\times 10^{-6}$ and $\xi_{\rm comp} = 10^{-4}$). 
The possibility of acceleration of particles in both shocks, either electrons or hadrons or both, can turn into a variety of radiation scenarios in which the high-energy $\gamma$-rays can be produced. As an example, we considered in a more detail two general scenarios for some specific values of the surface magnetic fields of the stars and acceleration efficiencies at the shocks created from the sides of the Eta Carinae star and the companion star. 
In the first one (model A), all high-energy emission comes from electrons accelerated in these two shocks. In the second model (model B), soft $\gamma$-ray emission comes from electrons and the higher energy $\gamma$-ray component is produced by hadrons. Both models can successfully describe the $\gamma$-ray  spectral features of the Eta Carinae (e.g., two-component spectrum, lack of strong variability at the multi-GeV energies). However, the leptonic model predicts a clear variability of the $\gamma$-ray emission above $\sim$100 GeV
with the period of the binary system. 
This variability is caused by the strong absorption of the produced $\gamma$-rays in the stellar radiation fields of both stars.
On the other hand, TeV $\gamma$-ray emission produced by hadrons in terms of the model B can either show variability in the case of their acceleration at the shock from the side of the Eta Carinae star ($\gamma$-rays from within the binary system) or not show variability if they are accelerated at the shock from the side of the  companion star ($\gamma$-rays produced within the larger scale nebula surrounding the binary system).

We have considered only a part of possible radiation scenarios
that could be expected in this type of a wind collision model. For other values of the surface magnetic field of the companion stars, locations of the shocks, or acceleration efficiencies, the acceleration of electrons and hadrons can be saturated at other energies. Therefore, a variety of spectral behaviours 
might be expected from other massive binaries. The general spectral features of the Eta Carinae could also be investigated with other parameters of the stars. Therefore, the set of parameters applied in our example modelling may not be unique. 

It is fairly difficult to conclude on the type of accelerated and radiating particles in the Eta Carinae binary system. In general, the leptonic model (model A) does not predict significant $\gamma$-ray emission above a few TeV
at any phase of the binary system. But the hadronic model (model B) predicts $\gamma$-ray emission at energies up to $\sim$10-100 TeV. Moreover, $\gamma$-ray spectra produced by hadrons injected into the Eta Carinae nebula should be independent of the phase of the binary system.
The most evident confirmation of the hadronic origin of $\gamma$-ray emission could come from  the detection of a neutrino signal from this source. Therefore,
we calculated the neutrino spectra expected for the model B and compared them  with the sensitivity of the 1 km$^2$ telescope. We showed that the neutrino spectra produced by hadrons within the nebula can be above the atmospheric neutrino background and within the sensitivity of the 1 km$^2$ telescope. However, the exact neutrino event rate in these telescopes is
difficult to predict because the energies of the injected hadrons depend on the phase of the binary system. The dependence of the injection rate of particles on the phase of the binary system cannot be reliably predicted at present.
Only future precise measurements of the $\gamma$-ray flux from the Eta Carinae as a function of the phase of this binary system may allow us to put limits on the injection rate of hadrons at different phases and thus allow us to estimate the expected neutrino event rates in large-scale neutrino telescopes
more precisely.

\begin{acknowledgements}
We would like to thank the anonymous referee for valuable comments.
This work is supported by the Polish MNiSzW grant N N203 390834 and NCBiR grant ERA-NET-ASPERA/01/10. Project co-funded by the European Union under the European Social Fund, POKL "Cz\l owiek – najlepsza inwestycja".
\end{acknowledgements}


\begin{thebibliography}{}

\bibitem{abdo10a} Abdo, A.A. et al. 2010 ApJ, 723, 649 
\bibitem{bed97} Bednarek, W. 1997 A\&A 322, 523
\bibitem{bed09} Bednarek, W. 2005 MNRAS 363, 46
\bibitem{br03} Benaglia, P., Romero, G.E. 2003 A\&A 399, 1121
\bibitem{bg70} Blumenthal, G.R., Gould, R.J. 1970 Rev.Mod.Phys. 42, 237
\bibitem{da08} Damineli, A., Hillier, D.J., Corcoran, M.F. et al. 2008 MNRAS 384, 1649
\bibitem{dh97} Davidson, K., Humphreys, R.M. 1997 ARA\&A 35, 1
\bibitem{eu93} Eichler, D., USO, V. 1993 ApJ 402, 271
\bibitem{hi01} Hillier, D.J., Davidson, K., Ishibashi, K., Gull, T. 2001 ApJ 553, 837
\bibitem{ks10} Kashi, A., Soker, N. 2010 ApJ, 723, 602
\bibitem{ley08} Leyder, J.-C., Walter, R., Rauw, G. 2008 A\&A 477, L29
\bibitem{ley10} Leyder, J.-C., Walter, R., Rauw, G. 2010 A\&A 524, 59
\bibitem{lip93} Lipari, P. 1993 APh 1, 195
\bibitem{mar10} Martin, J.C., Davidson, K., Humphreys, R.M., Mehner, A. 2010 AJ 139, 2056
\bibitem{meh10} Mehner, A., Davidson, K., Ferland, G.J., Humphreys, R.M. 2010 ApJ 710, 729
\bibitem{niel07} Nielsen, K.E., Corcoran, M.F., Gull, T.R., Hillier, D. J., Hamaguchi, K., Ivarsson, S., Lindler, D. J. 2007 ApJ 660, 669
\bibitem{ohd10} Ohm, S., Hinton, J.A., Domainko, W. 2010 ApJ 718, 1610
\bibitem{ob76} Orth, C.D., Buffington, A. 1976 ApJ 206, 312
\bibitem{par11} Parkin, E. R., Pittard, J. M., Corcoran, M. F., Hamaguchi, K. 2011 ApJ 726, 105
\bibitem{p09} Pittard, J.M. 2009 MNRAS 396, 1743
\bibitem{pc02} Pittard, J.M., Corcoran, M.F. 2002 A\&A 383, 636
\bibitem{pd06} Pittard, J.M., Dougherty, S.M. 2006 MNRAS 372, 801
\bibitem{rei06} Reimer, A., Pohl, M., Reimer, O. 2006 ApJ 644, 1118
\bibitem{sek09} Sekiguchi, A., Tsujimoto, M., Kitamoto, S., Ishida, M., Hamaguchi, K., Mori, H., Tsuboi, Y.~2009 PASJ 61, 629
\bibitem{sm03} Smith, N., Gehrz, R.D., Krautter, J. 1998 AJ 116, 1332
\bibitem{sm03} Smith, N., Gehrz, R.D., Hinz, P.M., Hoffmann, W.F., Hora, J.L., Mamajek, E.E., Meyer, M.R. 2003 AJ 125, 1458
\bibitem{tav09} Tavani, M., Sabatini, S., Pian, E. et al. 2009 ApJ 698, L142
\bibitem{um92} USO, V.V., Melrose, D.B. 1992 ApJ 395, 575
\bibitem{ver05} Verner, E., Bruhweiler, F. \& Gull, T. 2005 ApJ 624, 973
\bibitem{vio04} Viotti, R.F., Antonelli, L.A., Rossi, C., Rebecchi, S. 2004 A\&A 420, 527
\bibitem{wfl10} Walter, R., Farnier, C., Leyder, J.-C. 2010 A\&A, submitted (arXiv:1008.2533) 
\bibitem{ww87} Wdowczyk, J., Wolfendale, A.W. 1987 J.Phys. G 13, 411

\end{thebibliography}
\end{document}